\documentclass[11pt,twoside,A4]{article}
\usepackage{times,fancyhdr}
\usepackage[dvips]{graphicx}
\usepackage{latexsym}
\usepackage[affil-it]{authblk}
\usepackage{mathptm}
\usepackage{amsmath}
\usepackage{amssymb}
\usepackage{color}
\usepackage{setspace}
\usepackage{hyperref}
\usepackage[top=4cm, bottom=4cm, left=4cm, right=4cm]{geometry}

\def\beq{\begin{equation}}
\def\eeq{\end{equation}}
\def\beqn{\begin{eqnarray}}
\def\eeqn{\end{eqnarray}}

\newcommand{\be}{\begin{equation}}
\newcommand{\ee}{\end{equation}}
\newcommand{\bea}{\begin{eqnarray}}
\newcommand{\eea}{\end{eqnarray}}

\begin{document}

\title{General Relativity with Local Space-time Defects}
\author{Sabine Hossenfelder, Ricardo Gallego Torrom\'e}
\affil{\small Frankfurt Institute for Advanced Studies\\
Ruth-Moufang-Str. 1,
D-60438 Frankfurt am Main, Germany
}
\date{}
\maketitle
\begin{abstract}
General relativity is incomplete because it cannot describe
quantum effects of space-time. The complete theory of quantum gravity
is not yet known and to date no observational evidence exists that
space-time is quantized. However, in most approaches to quantum gravity the space-time
manifold of general relativity is only an effective limit that, among other things like
higher curvature terms, should receive corrections stemming from
space-time defects. We here develop a modification of
general relativity that describes local space-time defects and solve the
Friedmann equations. From this, we obtain the time-dependence of the average
density of defects. It turns out that the defects' average density dilutes quickly, somewhat faster
even 
than radiation. 

\end{abstract}

\section{Introduction}

The currently accepted theory for gravity -- general relativity -- is not compatible with
quantum field theory, the conceptual framework which the standard model of particle physics builds on.
Therefore, general relativity is  today understood as an effective theory that is
only approximately correct. At high energies it needs to be completed, in a mathematically
consistent way, both to render it renormalizable and to couple it to quantum fields.
While solving this problem does not mean that gravity necessarily must be quantized, we will here -- as is common in the literature --
refer to the sought-after UV-completion as `quantum gravity.'

While the effects of quantum gravity are expected to be strong only in regimes where
the curvature is close by the Planck scale, not all deviations from general relativity are 
relevant only at high energies. Symmetry
violations in particular are known to impact low energy physics even if they originate
in the ultra-violet, amply demonstrated for example by Lorentz-invariance violation \cite{Mattingly:2005re}.
In general
relativity, one assumes that space-time is described by a differentiable manifold
giving rise to local conservation laws. If the underlying theory of
quantum gravity does not respect this symmetry -- and there is no reason why it
should -- then local conservation laws can be violated. Indeed, one should
generically expect this to be the case.

In absence of a fully-fledged theory for quantum gravity, one cannot derive observable
consequences. One can, however, quantify them by help of phenomenological models.
In \cite{Hossenfelder:2013yda,Hossenfelder:2013zda}, a framework for space-time
defects was developed that respects global Lorentz-invariance. In this framework,
local space-time defects couple to particles through random kicks that change the
particle's momentum which is mathematically encoded in a stochastic contribution
to the derivative operator.

Space-time defects have been discussed in the literature for some time \cite{Klinkhamer:2003ec,Schreck:2012pf,Klinkhamer:2017nhl}. Older
models are not (locally) Lorentz-invariant and are now in tension
with data. We will not consider these here. Some newer models respect Lorentz-invariance but rely on additional
exchange fields to couple Standard Model particles to the defects. We will not consider this case here
either. Instead, we will in this paper extend
the approach proposed in \cite{Hossenfelder:2013yda,Hossenfelder:2013zda}
which respects local Lorentz-invariance and does not necessitate additional fields.

In the following we will generalize the previously developed
phenomenological model for local space-time defects to curved backgrounds.
As we will show, this implies a modification of Einstein's field equations by a change of the
covariant derivative. We will then go on to derive the field equations for the defects' average
density. Knowing the time-dependence of the defect-density with the expansion of
the universe is relevant input to understanding possible phenomenological
consequences.

Throughout this paper we use units in which $c=\hbar=1$. The signature
of the metric is $(1,-1,-1,-1)$, and the dimension of the manifold is $4$. Small
Greek indices run from $0$ to $3$. Bold-faced quantities denote tensors
whose coordinate components are given by the respective symbol with
indices. Eg, ${\bf p}$ denotes the vector with components $p_\mu$, ${\bf g}$
is the two-tensor with entries $g_{\kappa \nu}$ and so on.

\section{Local Defects in Flat Space}

For the benefit of the reader, we will first briefly summarize the model for
local space-time defects developed in \cite{Hossenfelder:2013zda}.

We start
with the assumption of Poincar\'e-invariance. For the distribution of defects
in space-time we use the (only known) stochastic distribution that is
-- on the average -- both homogeneous and Lorentz-invariant \cite{Bombelli:2006nm}. It is a
Poission-distribution according to which the probability to find $N$ defects in a four-volume $V$ of spacetime is
\begin{align}
P_N(V)=\, \frac{(\beta V)^N\,\exp(-\beta V)}{N!}~.
\label{probability of sprinkling N points}
\end{align}

We then have to parameterize what happens at a defect. It was assumed in \cite{Hossenfelder:2013zda},
that defects induce a violation of energy-momentum conservation because they represent
a deviation from the smooth structure of the underlying manifold. A particle (or wave) with incoming
momentum ${\bf p}$ will scatter on the defect and exit with momentum ${\bf p}'$.
The difference between the two momenta, ${\bf k} := {\bf p}' - {\bf p}$, can formally be assigned to
the defect. However, we want to emphasize that assigning this momentum to the defect is merely for book-keeping
and does not mean
the defect actually carries a momentum in any physically meaningful way.

The change of momentum that happens at a defect is assumed to be stochastically distributed. The requirement
of Lorentz-invariance then implies that this distribution can only be a function of the
three invariants:
\beqn
p_\nu p^\nu =  m^2~,~k_\nu p^\nu =  M^2~,~ k_\nu k^\nu = a^2 M^2~,
\eeqn
where $m$ is the mass of the incident particle (and may be equal to zero), $M$ is a parameter of dimension mass, and
$a$ is a dimensionless constant expected to be of order one.

Importantly, the direction of the outgoing momentum ${\bf p}'$ (or its distribution,
respectively) is a function of
the ingoing momentum. For this reason, the defects do not introduce a preferred
frame. While the scattering at a defect has a preferred direction, this direction is
entirely determined by the incoming particle. It is in this exact way that the model preserves local Lorentz-invariance: It does
not introduce a fundamental preferred frame. To the extent that a frame is preferred,
this frame is -- as usual -- defined by the dynamics of matter fields.

Finally, the coupling of matter to the defects is made by replacing the
usual partial derivative, $\partial_\nu$, with $\partial_\nu + W_\nu$, where
$W_\nu$ is a vector-valued random variable, defined on the point-set of defects,
with the probability distribution of $k_\nu$ at each point.

With these ingredients, one can write down a modified Lagrangian for matter
coupled to the defects and calculate cross-sections.

As was shown in \cite{Hossenfelder:2013zda}, the observational consequences of
defects become more pronounced the smaller the energy of the particle that is
being scattered, and the longer its travel time. To be more precise, what matters is not the travel-time but the world-volume
swept out by the wave-function --  a direct consequence of requiring
Poincar\'e-invariance,

This means that the best constraints on defects come from cosmological data. However,
to analyze cosmological data, it is necessary to deal with an expanding background.
We will therefore here further develop the model so that we can deal with
curved space-time and study Friedmann-Robertson-Walker cosmologies in
particular. The most important question we would like to address is how the average
density of space-time defects changes with time.

\section{Local Defects in Curved Space}

We now turn towards the main purpose of the paper, the question how to generalize
the model for local space-time defects to a curved background.

\subsection{The Connection}

In flat space, we coupled quantum fields to defects by adding
localized, stochastic contributions to the partial derivative. This is
straight-forward to generalize to curved space-time by instead adding
these contributions to the covariant derivative. For this purpose,
we define a new derivative
\beqn
\widetilde { \nabla} := {\nabla} + {\bf Q}~,
\eeqn
where ${\nabla}$ is the usual Levi-Civita-connection, ie the unique
connection that is both metric-compatible
and torsion-free.

We want  to
inflict a minimum of harm on general relativity and hence require
that the new derivative, $\widetilde \nabla$, is generally covariant. This
means that the additional term ${\bf Q}$ is (unlike the
Christoffel-symbols) a three-tensor. The new derivative, however, is
no longer the usual Levi-Civita-connection.

Since torsion has no effect on the geodesic equation, we will in the following
not take it into account. (See, however, the discussion in section \ref{disc}). We will hence assume
that $\widetilde \nabla$ is torsion-free, ie that
\begin{align}
\widetilde{\nabla}_X Y-\widetilde{\nabla}_Y X-\,[X,Y]=0~,
\label{torsionfreecondition 1}
\end{align}
for arbitrary vector fields $X,Y$.
In coordinate notation this means that the connection coefficients $\widetilde \Gamma$ that belong to
$\widetilde \nabla$ are of the form
\beqn
\widetilde\Gamma^\kappa_{\;\nu\mu}  = \Gamma^\kappa_{\;\nu\mu} + Q^\kappa_{\;\nu\mu}~,
\label{relation Gamma tilde Gamma}
\eeqn
where
${\bf Q}$ must be symmetric in the lower two indices
\beqn
Q^\kappa_{\;\;\nu\mu} = Q^\kappa_{\;\;\mu\nu}~.
\eeqn
The non-metricity tensor ${\bf Q}$ can be expressed as
\beqn
g_{\sigma \tau} Q^{\tau}_{\; \;\mu \nu} = - \frac{1}{2} \left( \widetilde \nabla_\mu g_{\nu \sigma} + \widetilde \nabla_\nu g_{\mu \sigma}
- \widetilde \nabla_\sigma g_{\mu \nu} \right) ~. \label{qcyc}
\eeqn

Similar to the case of flat space, we will then assume that a defect imparts
a stochastic kick on an incoming particle and that the kick's distribution (though not its value) is entirely determined by the outgoing
momentum. We will hence assume that ${\bf Q}$ is proportional to the vector-valued
random variable
${\bf W}$ that has support on the set of defects. Together with torsion-freeness and the index structure Eq.~(\ref{qcyc}),
this means ${\bf Q}$ must be of the form
\beqn
Q^\rho_{\;\;\mu \nu} =
\,\frac{1}{2}\left(W_\mu\,\delta^\rho_{\;\nu}+
\,W_\nu\,\delta^\rho_{\;\mu}-\,W^\rho g_{\mu\nu}\right)~,\label{QW}
\eeqn
where we have absorbed a possible pre-factor into ${\bf W}$.

 Such a modification of the covariant derivative has an interesting physical interpretation,
 which is a non-conservation of the volume element, $\sqrt{-g}$,
 where $g=\rm{det} ( {\bf g}) $. To see this, recall that usually $\nabla g = 0$ and
 note that the local relation
\begin{align}
\partial_\lambda \ln (-g)=\,\frac{1}{g}\,\partial_\lambda g=\,g^{\mu\sigma}\,\partial_\lambda\,g_{\sigma\mu}
\label{partical derivative of the determinant}
\end{align}
is fulfilled regardless of what the connection is.

With use of Eq.~\eqref{partical derivative of the determinant}, we can calculate the
new covariant derivative of the volume element:
\beqn
\frac{1}{-g}\,\widetilde\nabla_\lambda (-g) & =& g^{\mu\sigma}\,\widetilde\nabla_\lambda g_{\sigma\mu} \nonumber \\
& =&g^{\mu\sigma}\,\left(\nabla_\lambda g_{\sigma\mu} - Q^\rho_{\; \;\lambda\sigma}g_{\mu\rho}-Q^\rho_{\; \;\lambda\mu}g_{\rho\sigma}\right) \nonumber \\
& =& - 2 Q^\rho_{\; \; \lambda\rho}~,
\eeqn
where we have used that $\nabla_\lambda g_{\sigma\rho}=0$.
This can be rewritten as
\beqn
\frac{1}{{-g}} \widetilde\nabla_\lambda( {-g}) = - 2 W_\lambda~. \label{volel}
\eeqn

We hence see that this connection's vector-valued non-preservation
of the volume-element is a natural way to describe space-time defects because it induces a violation of energy-momentum conservation, an effect that comes with space-time defects \cite{Hossenfelder:2013yda,Hossenfelder:2013zda}. 

Indeed, it was
demonstrated in  \cite{tHooft:2008mxs,Arzano:2014nwa,Wieland:2016exy,vandeMeent:2011wr} that conical
space-time singularities have properties similar to the ones associated with space-time
defect: a) space-time is flat except for one point, b) at that one point the curvature is divergent and hence
ill-defined,
and c) passing by near the singularity/defect imparts a momentum on the particle that can
be expressed as a locally acting Lorentz-boost.

While the conical singularities are examples the reader might
want to keep in mind, we would like to emphasize that the defects we consider here differ
from conical singularities in that they do not have a fixed orientation, but rather a distribution
over orientations that depend on the momentum of the incident particle.

The connection $\widetilde{\nabla}$ can also be expressed as
\beqn
\widetilde{\nabla}_\rho g_{\mu\nu}= W_\rho g_{\mu\nu}~,
\label{projective metric compatibility}
\eeqn
which has previously been discussed in the literature under the name `projective metric compatibility' or `vector non-metricity' \cite{HehlLordSmalley, Gasperini,Stelmach}. 

The new derivative $\widetilde \nabla$ has a curvature-tensor associated
to it, which is as usual (following the convention of \cite{Wald}) defined by the commutator of the covariant derivatives of
an arbitrary $1$-form ${\bf A}$:
\beqn
[ \widetilde \nabla_\sigma, \widetilde \nabla_\kappa] A_\alpha := \widetilde R^{\mu}_{\; \; \sigma \kappa \alpha} A_{\mu}~.
\eeqn
From this curvature tensor, we can construct the curvature scalar,
which will serve as the Lagrangian for our modified theory of gravity.

 \subsection{The Lagrangian for Gravity Coupled to Defects}

 Since we have an additional vector field that describes the covariant
derivative, we will use the Palatini-formalism to derive the equations of
motion. 
In the Palatini-formalism, one makes an independent variation over the metric
and the connection separately.
If one uses the Einstein-Hilbert action (ie, the curvature scalar) and assumes that the
connection is torsion-free (as we have done), then the additional equations one obtains in the
Palatini-formalism require the connection to also be metric-compatible.

One may think this
is because in the Palatini-formalism the Einstein-Hilbert action is no longer the unique
choice since there are various other terms that can be constructed from the connection,
for example those composed of covariant derivatives of the volume-element. But interestingly,
as was shown in \cite{Burton-Mann}, even with the additional terms added,
the Palatini-formalism gives back General Relativity under quite general circumstances.

However, a central assumption for the conclusion in \cite{Burton-Mann} is that the Lagrangian
of the matter fields does not make a contribution to the constraint equations for the
connection that are derived from the Palatini-formalism. This, however, will in general
not be the case. While a gauge-field effectively only cares about the partial derivative
so long as the connection is torsion-free because the field-strength tensor is
anti-symmetric, this is not the case for fermion fields.

We will hence use the formalism of \cite{Burton-Mann}, but add matter sources. For simplicity, we
will restrict the analysis presented here to a single Dirac field with mass $m$,  though our approach 
can easily be extended to more general cases.

For a Dirac field, the covariant derivative is defined by help of the spin connection (see for instance \cite{Pollock}). It can be constructed from the tetrad, $e^a_{\; \mu}$, and the connection, $\widetilde{\nabla}$, by the relation
\begin{align}
\widetilde w^{ab}_{~~~\mu} := \,e_{\nu}^{\; \; a} 
\widetilde{\Gamma}^\nu_{\;\;\sigma \mu}\,e^{\sigma b}+ e_{\nu}^{\; \; a} \partial_\mu e^{\nu b}~.
\label{spin connection 1}
\end{align}
This spin connection acts on  sections of the bundle of Dirac spinors and determines the covariant derivative operator by
\begin{align}
\widetilde D_\mu= \partial_\mu-\frac{\rm i}{4}\widetilde w^{ab}_{~~~\mu}\,[\gamma_a,\gamma_b]~.
\label{spin derivative}
\end{align}
Using Eqs.~\eqref{relation Gamma tilde Gamma} and (\ref{QW}) this can be expressed in terms of ${\bf W}$ as:
\beqn
\widetilde{D}_\mu   = D_\mu -\frac{\rm i}{4}\,e_{\mu}^{\; \; a}\,e^{\sigma b}W_\sigma \,[\gamma_a,\gamma_b]~,
\eeqn
where $D_\mu$ is the spin connection associated with the usual Levi-Civita connection, $\nabla$.
The  Lagrangian for the Dirac field is then
\beqn
{\cal L}_{\rm M} = 
\frac{\rm i}{2} \left(\overline{\psi}  \gamma^{\; \mu} D_\mu \psi -  \overline{D_\mu \psi} \gamma^{\; \mu} \psi \right) -m\overline{\psi}\psi+ \frac{1}{2} e_\mu^{\;\;a} e^{\sigma b} W_\sigma \overline{\psi} [\gamma_a,\gamma_b]\,\gamma^{\;\mu} \psi ~.
\eeqn

The generalization of the ansatz from \cite{Burton-Mann} with the addition of a Dirac field therefore starts
with the action
 \beqn
 S[g, \widetilde \Gamma, \psi ] &=& \int {\rm d}^4 x \sqrt{-g} \big(  \frac{1}{16\pi G} \big(\widetilde{R} + 
 C_1(\widetilde{\nabla}_\nu g^{\mu\rho})(\widetilde{\nabla}^\nu g_{\mu\rho})+ C_2  V_\rho V^\rho \nonumber \\
 && + C_3(\widetilde{\nabla}_\rho g_{\mu\nu})(\widetilde{\nabla}^\mu g^{\rho \nu})+ C_4 V_\rho Z^\rho+ C_5 Z_\rho Z^\rho \big)+ \cal{L}_{\rm M} \big)~, \label{BMaction}
 \eeqn
where $C_1,C_2,C_3, C_4, C_5$ are dimensionless constants and
\begin{align}
V_\rho=\,\frac{1}{\sqrt{- g}} \widetilde{\nabla}_\rho \sqrt{-g}\quad,\quad Z^\mu=\widetilde{\nabla}_\rho g^{\rho\mu}~. \label{this1}
\end{align}
For the case of vector non-metricity, the two vectors ${\bf V}$ and ${\bf Z}$ from \cite{Burton-Mann} are:
\beqn
V_\nu = - W_\nu ~,~ Z_\nu=  - W_\nu~. \label{ZandV}
\eeqn
By use of the relations (\ref{this1}) and (\ref{ZandV}) one convinces oneself that the additional terms
in the action Eq.\ (\ref{BMaction}) are all propotional to each other. The five different constants therefore
can be replaced with merely one constant that is a linear combination of $C_1$ to $C_5$.

One obtains the field equations from variation of the action with respect to the metric and then inserting the relations (\ref{ZandV}). This results in
\beqn
{R}_{\mu\nu}-\frac{1}{2}R g_{\mu\nu}-2\left(\nabla_\mu W_\nu+\nabla_\nu W_\mu\right)+2 g_{\mu\nu}\nabla_\alpha W^\alpha+ (8+C)W_\mu W_\nu+ 4g_{\mu\nu}W^2= 8\pi G T_{\mu\nu}~,
\label{general field equations}
\eeqn
where $C$ is a dimensionless constant that is the (not so relevant) linear combination of the constants in Eq.~(\ref{BMaction}).
In (\ref{general field equations}) the Ricci tensor, $R_{\mu\nu}$, and the scalar curvature, $R$, are the ones associated with the Levi-Civita connection, and the stress-energy tensor is, as usual, defined by
\beqn
T_{\mu\nu} := - 2 \frac{\delta {\cal L}_{\rm M}}{\delta g_{\mu\nu}} + g_{\mu \nu} {\cal L}_{\rm M}~.
\eeqn
The field equations obtained this way are identical to the ones derived in \cite{Gasperini,Stelmach}.

The Dirac field $\psi$ satisfies the equation
\beqn
{\rm i}\gamma^{\; \mu} D_\mu\,{\psi}+  \frac{1}{2}\,e_\mu^{\;\;a}\,e^{\sigma b} W_\sigma [\gamma_a,\gamma_b]\,\gamma^{\; \mu} \psi  -m\psi=0~,
\label{Fermion equation 1}
\eeqn
and the conjugate field obeys the respective conjugated equation. 

The Euler-Lagrange equation for ${\widetilde \Gamma}$ together with (\ref{ZandV}) leads to the relation
\beqn
\frac{1}{2} e^a_{\;\;\mu} e^{\sigma b} \overline{\psi}\gamma^{\; \mu} [\gamma_a,\gamma_b]\psi + 2 C \frac{1}{16\pi G} W^\sigma=0~. \label{Weq}
\eeqn
One sees clearly that in the absence of the matter field, this would merely lead to the conclusion that
${\bf W} \equiv 0$, so that we would be returned to normal general relativity. This is the
conclusion drawn in \cite{Burton-Mann}. However, in the
presence of matter fields, this is not necessarily so; in this case, ${\bf W}$ may be non-vanishing.

With some algebraic manipulations, Eq.~(\ref{Weq}) can be simplified to  
\begin{align}
C W_\nu = - 24 \pi G J_\nu~ ,
\label{equationforW}
\end{align}
where $J_\nu=\overline{\psi}\,\gamma_\nu \psi$ is the vector current.
This relation, most importantly, implies that ${\bf W}$ is  entirely determined
by the matter fields. 

We therefore see that the presence of space-time defects
induces an order six operator, suppressed by the (square of the) Planck mass, that effectively gives rise to a four-fermion coupling.
The vector-field ${\bf W}$ can be removed from the matter-field's equation
of motion Eq.~\eqref{Fermion equation 1} which gives 
\begin{align}
{\rm i} \gamma^{\; \mu} D_\mu 
{\psi}+ 12\frac{1}{C} \pi G (\overline{\psi} \gamma_\mu \psi) \gamma^{\; \mu} \psi- m\psi=0~
\label{Fermion equation 2}
\end{align}
and its hermitian conjugate. 

We note that this result bears similarity to the {\sc BCS} condensates discussed in \cite{Alexander:2008vt,Poplawski:2011wj}, where a four-fermion interaction was induced by torsion. However,  torsion leads to a coupling with the axial current, whereas we have a coupling with the vector current that stems directly from the vector which quantifies the non-metricity.

\subsection{Conservation of the current}

In the action Eq.~(\ref{BMaction}), no derivatives acting on ${\bf W}$ appeared. That is fortunate because 
we had defined ${\bf W}$ only on a discrete set of points. The observant reader will have noticed, however,
that to derive the field equations Eqs.~(\ref{general field equations}), we have assumed that
the field is differentiable in order to make sense of derivatives acting on it.  We have allowed ourselves
this freedom because, in the next section on cosmology, we will deal
with the field's expectation value rather than the random variable itself. In this case, then, it is meaningful to
speak about derivatives. For this reason, we will here also briefly look at the conservation
laws that are obeyed on the average.

The equation of motion for the Dirac field \eqref{Fermion equation 2} and its hermitian conjugate imply a conservation
law that is analogous to the usual conservation of the current. We have not included gauge-fields here,  but we still
have a conservation law stemming from the global U$(1)$ symmetry. From the modified Dirac equation we obtain
\beqn
\nabla_\mu W^\mu & =& 24 
\pi G \frac{1}{C}\left(D_\mu(\overline{ \psi})\gamma^{\; \mu} \psi+ \overline{\psi}\gamma^{\; \mu} D_\mu\psi\right)= 0~ .\label{conservation law}
\eeqn
where we have used  $D_\mu \gamma_{\rho}=0$ \cite{Pollock}.

\subsection{Bianchi identities and violation of stress-energy conservation}

Next, we will look at the conservation of the stress-energy tensor which, in general relativity, is a direct consequence
of the Bianchi-identities. We expect this conservation law to be modified, but also that it must be possible to
construct a new, modified, conservation law. 

For torsion-free connections, the second Bianchi identities can be expressed in local coordinates as \cite{Kobayashi-Nomizu,Wald}
\beqn
\widetilde{\nabla}_\lambda \widetilde{R}^{\nu}_{\; \rho\sigma \mu} + \widetilde{\nabla}_\rho \widetilde{R}^{\nu}_{\;\sigma \lambda \mu} + \widetilde{\nabla}_\sigma \widetilde{R}^{\nu}_{\; \lambda \rho \mu} = 0 ~.
\label{second Biachi identity}
\eeqn
This relation is valid for any torsion-free affine connection, like the one we are using here (see for example Theorem 5.3. in \cite{Kobayashi-Nomizu}).
The contracted Bianchi identities are obtained by first taking the trace of the above expression. This results in

\beqn
\widetilde{\nabla}_\lambda \widetilde{R}^{\lambda}_{\;\;\rho\sigma \mu} + \widetilde{\nabla}_\rho \widetilde{R}_{\sigma\mu} - \widetilde{\nabla}_\sigma \widetilde{R}_{\rho \mu} = 0 ~, \label{step1}
\eeqn
where, as usual, the Ricci-tensor is defined by $\widetilde R_{\mu\nu} = \widetilde R^\kappa_{\;\;\mu \kappa \nu}$, and
we have used the cyclicity of the curvature-tensor
\beqn
R^\nu_{\; [   \kappa \mu \lambda ]} = 0~, 
\eeqn
which also holds in the absence of torsion.

Next, we contract Eq.~(\ref{step1})  with  $g^{\rho \mu}$ and obtain
\beqn
\widetilde{\nabla}_\lambda (\widetilde{R}^{\lambda}_{\;\; \rho\sigma \mu}g^{\rho\mu}) 
+ \widetilde{\nabla}_\rho (\widetilde{R}^\lambda_{\; \;\sigma\lambda\mu}g^{\rho\mu}) - 
\widetilde{\nabla}_\sigma \widetilde{R} = -( \widetilde{R}^{\lambda}_{\; \;\rho \sigma\mu}\widetilde{\nabla}_\lambda g^{\rho\mu}+\,\widetilde{R}^{\lambda}_{\; \; \sigma\lambda\mu} \widetilde{\nabla}_\rho g^{\rho\mu}- \widetilde{R}_{\rho \mu}\,\widetilde{\nabla}_\sigma g^{\rho\mu})~.
\eeqn
By using the relation
\beqn
\widetilde{\nabla}_\lambda g^{\rho\mu}= -W_\lambda g^{\rho\mu}~,
\eeqn
the contracted Bianchi identities can be expressed in terms of ${\bf W}$ as
\beqn
\widetilde{\nabla}_\lambda \widetilde{R}^{\lambda}_{\; \; \rho\sigma}{}^{\rho}  + \widetilde{\nabla}_\rho \widetilde{R}^\lambda_{\; \; \sigma\lambda}{}^\rho  -\, \widetilde{\nabla}_\sigma \widetilde{R}=
W_\lambda \widetilde{R}^{\lambda}_{\; \;\rho\sigma}{}^\rho + W_\rho \widetilde{R}_{\sigma}^{\;\;\rho}-  W_\sigma \widetilde{R}~. \label{conslaw1}
\eeqn

The terms on the left side of Eq.~(\ref{conslaw1}) can be written as 
$2 \widetilde \nabla_\lambda \widetilde G^\lambda_{\;\; \sigma}$
for a generalized Einstein-tensor
\beqn
2 \widetilde G_{\mu \nu} = \widetilde{R}_{\mu \rho\nu}{}^{\rho}  +  \widetilde{R}^\rho_{\;\; \nu\rho \mu}  -  g_{\mu \nu} \widetilde{R}~.
\eeqn
The field equations obtained by a variation of $g_{\mu\nu}$ in the action are of the form
\beqn
\widetilde{G}_{\lambda\sigma} + CW_\lambda  W_\sigma=  8\pi G T_{\lambda\sigma}.
\label{Einstein equations 2}
\eeqn
This, in combination with the contracted Bianchi identity, finally leads to the new conservation law
\beqn
8 \pi G \widetilde{\nabla}^\mu T_{\mu\nu} = W^\mu \widetilde{G}_{\mu\nu}+C\,\widetilde{\nabla}^\mu (W_\mu W_\nu)~. \label{setconsnew}
\eeqn

Alternatively, we can take the ${\bf W}$-dependent terms that stem from the curvature in Eq.~(\ref{general field equations}) and
assign them a new tensor
\beqn
\tau_{\mu \nu} := 2\left(\nabla_\mu W_\nu+\nabla_\nu W_\mu\right)-2 g_{\mu\nu}\nabla_\alpha W^\alpha- (8+C)W_\mu W_\nu- 4g_{\mu\nu}W^2 ~. \label{tau}
\eeqn
The interpretation of this tensor is the stress-energy associated with the defects. The sum of this tensor and the usual
stress-energy-tensor then obeys the normal conservation law
\beqn
 {\nabla}^\mu \left( 8 \pi G T_{\mu\nu} + \tau_{\mu \nu} \right) = 0~, 
\eeqn
but generically neither term is separately conserved.

\section{Friedmann-Robertson-Walker spacetimes with defects}

In this section we will look at cosmology with space-time defects. To that end, we will assume
isotropy and homogeneity are fulfilled on the average. In particular, we will promote ${\bf W}$ from
being a random variable defined only on a set of points to a differentiable vector field. The vector
field should not be interpreted as encoding the number-density of defects. It encodes the average
energy and momentum that is transferred by the defects. From this field, we will derive the 
average stress-energy tensor associated with the presence of the defects
which acts as a source-term for the field equations. This, then, will allow us to calculate
the time-dependence of the field itself by solving the field-equations.

The distance scale at which this approximation should be appropriate is for space-time
volumes in which
there is a large number of defects, but that is still much smaller than the fourth power of
the curvature radius, ie $\ll 1/\Lambda^{2}$, where $\Lambda$ is the cosmological constant. 
Since the typical density of nodes in a space-time that is fundamentally made
from a network should be set by the Planck-scale, there are many orders of
magnitude in which this limit is good. Indeed, in
\cite{Hossenfelder:2013yda,Hossenfelder:2013zda}, it was found that constraints
from Minkowski space limit the density merely to be smaller than about an inverse femtometer to
the fourth power.

It is not a priori clear whether the use of average values to describe the universe
on cosmological scales is justified 
because the field equations of gravity are non-linear. This is a well-known problem in
general relativity
and we do not have anything new to say about it. For details the reader may refer to the review \cite{Bolejko:2016qku}.
We will here, as common in the literature,
simply work with the averages in the hope that this debate will be
resolved at some point. 

\subsection{Derivation of Friedmann-equations}

We start with the ansatz for a Friedmann-Robertson-Walker ({\sc FRW}) metric 
\beqn
{\rm d}s^2 =  {\rm d}t^2 - a(t)^2 \left( \frac{{\rm d}r^2}{1-kr^2}  + {\rm d} \Omega^2 \right)~,
\label{Robertson-Walker}
\eeqn
where $k \in \{-1,0.1 \}$ is the constant curvature on spatial hypersurfaces. For ${\bf W}$, we make the ansatz $(Y(t),0,0,0)$, and for the stress-energy tensor we assume the common form of a perfect fluid
\beqn
T^\mu_{\;\; \nu}= {\rm{diag}} (\rho, -p,-p,-p)~.
\label{fluid tensor}
\eeqn
One could make a more general ansatz for ${\bf W}$ in which the spatial components do not vanish, but one would find
later that the field equations demand they do vanish because such components would induce off-diagonal entries. 

Now, onto the field equations. The off-diagonal equations are automatically fulfilled since ${\bf W}$ only has a
 zero-component which is a function of $t$ alone.  The first and second Friedmann-equation read:
\beqn
\left( \frac{\dot{a}}{a} \right)^2 &=& \frac{8}{3}\pi G \rho  - \frac{k}{a^2}  - 2 \frac{\dot{a}}{a} Y + \frac{2}{3} \dot Y - (4+C/3) Y^2~, \label{Friedmanneq1}\\
\frac{\ddot{a}}{a} &=& -4\pi G\left(p+\frac{\rho}{3}\right) - \frac{4}{3} \dot Y + \frac{C}{6} Y^2 ~.
\label{Friedmanneq2}
\eeqn

The reader will note immediately that these equations differ from the usual Friedmann-equations not only
by the additional sources, but by the relation between the sources. The
reason is that the sources do not fulfill the usual conservation law, but the modified one Eq.~(\ref{setconsnew}). Since
the violation of (the usual) stress-energy conservation was the point of our exercise, let us make this
important consistency check explicitly. For the isotropic and homogeneous ansatz, the zero-component of the
conservation law reads:
 \beqn
 4 \pi G  \left( \dot{\rho}+ 3 \frac{\dot{a}}{a} (\rho + p )  \right)  = (C+12) Y \dot Y + \ddot Y + 3 \left( \frac{\ddot a}{a} + 
\left(\frac{\dot a}{a} \right)^2 \right) Y + \frac{3(8+C)}{2} \frac{\dot a}{a} Y^2  ~.
 \eeqn
(The other equations vanish identically.)
One confirms easily that when one takes the time-derivative of the first Friedmann-equation and inserts this
modified conservation law, one obtains -- correctly -- the second Friedmann-equation.
 
To solve the modified Friedmann-equations Eqs.~(\ref{Friedmanneq1}) and (\ref{Friedmanneq2}), we draw upon  the equation which we derived for the conserved current, Eq.~\eqref{conservation law}.
For the {\sc FRW} case, this leads to the simple relation
\beqn
\dot{Y}= - 3 Y \frac{\dot a}{a}~.
\label{conservation law 2}
\eeqn
which has the solution
\beqn
Y(t) = \frac{Y_0}{a(t)^3}~, \label{Ysol}
\eeqn
where $Y_0$ is some initial value. 

Next, we  insert this expression into the first Friedmann-equation (\ref{Friedmanneq1}) which gives
\beqn
\left( \frac{\dot{a}}{a} \right)^2 
&=& \frac{8}{3}\pi G \rho  - \frac{k}{a^2}  - 4 Y_0 \frac{\dot a}{a^4} - (4+C/3) \frac{Y_0}{a^6}~. \label{adota}
\eeqn
This equation can now be solved for $\rho$, and the solution, together with Eq.~(\ref{Ysol}), can be inserted
into the second Friedmann equation (\ref{Friedmanneq2}). This decouples the system. The resulting
equations
\beqn
8 \pi G \rho(t) &=& 3 \left( \frac{\dot a}{a} \right)^2 - (C+12) \frac{Y_0^2 }{a^6} - 12 Y_0 \frac{\dot a}{a^4}~ + \frac{3k}{a^2}~,
\label{rhoeq}\\
\frac{\ddot{a}}{a} &=& - \frac{1}{2} \left( \frac{\dot a}{a} \right)^2  -  \frac{k}{2 a^2} -2 \frac{Y_0^2 }{a^6} + 2 Y_0 \frac{\dot a}{a^4}~\label{ddota}.
\eeqn
can be integrated numerically, which we will do in the next section.

\subsection{Results}

In this subsection we display the results of a numerical integration of Eq.~(\ref{adota}). In the
previous subsection we derived the general equations valid for any equation of state, but here we
will examine in particular  a flat and matter-dominated universe, ie we will set $k=0$ and $p=0$. 
The latter choice means that we neglect the presence of radiation. 

Our solutions
will then depend on various parameters. First, there are the initial values for the scale-factor and the energy-density,
$a_0$ and $\rho_0$. While these affect the quantitative result, they are not relevant for the scaling
of the solution with $t$. We moreover have the constant $C$ that determines how strongly
the additional terms contribute, and the initial value $Y_0$ for $Y(t)$ from Eq.~(\ref{Ysol}) . 

Let us begin with some general considerations. The second term on the right side of Eq.(\ref{rhoeq}) and
the second term on the right side of Eq.~(\ref{ddota}) go with
$1/a^{6}$ and therefore will become irrelevant compared with the other terms quickly. The leading
deviation from the usual case therefore goes with $\dot a/a^4$. Assuming that $\dot a$  as
usual decreases, these contributions too are subdominant to the usual term. In this case,
we can expect that approximately $a \dot a^2 \sim$ constant and the correction terms scale
as $a^{-9/2}$.

For the integration we will use
initial conditions so that at the present time, $t_0$, we have $8 \pi G \rho_0 = 1$ and $a_0 = 1$, ie the density and scale factor are 
measured in relative to today's density and scale factor.

Because of the above mentioned scaling considerations, we further use initial conditions with
$Y_0^2 \ll 8 \pi G \rho_0$. This is justified if we assume that both -- the energy-density of matter and that in the defects --  started out about the 
same value at some early time, say, at the Planck time. Relative to the baryonic density, the
leading term in $00$-component of the stress-energy of the defects Eq.~(\ref{tau}) will get an additional drop from $\dot a/a 
\sim \sqrt{\Lambda}/m_{\rm p}$, where $\Lambda$ is the cosmological constant and $m_{\rm p}$ is the Planck scale. This
means if $Y$ started with a value $Y_{\rm p} \sim m_{\rm p}$ at Planckian times, then today it will be at $Y_0 \sim \sqrt{\Lambda}^3/m_{\rm p}^2$.  A realistic value would be $Y_0 \sim 10^{-29}$,
but that would make a numerical treatment infeasible. Instead, we will use more manageable values of $\sim 10^{-2}$ 
that illustrate the behavior more clearly.

\begin{figure}[th]
\centering
\vspace*{-1.3cm}\hspace*{0.5cm}
\includegraphics[width=10cm]{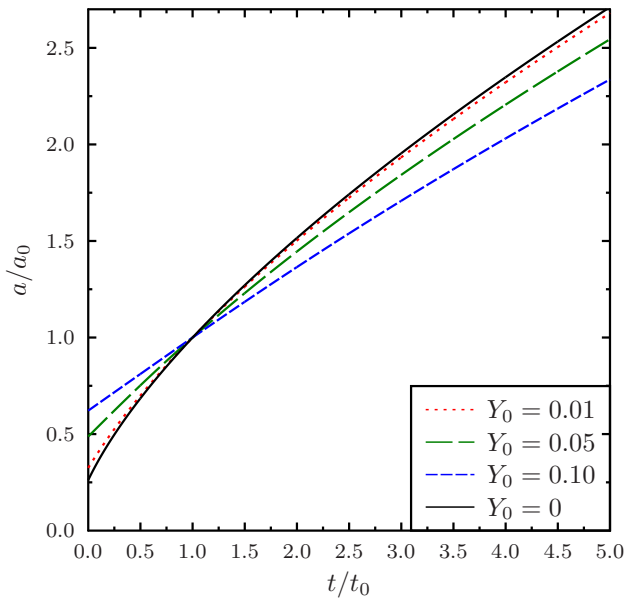}
\vspace*{-1cm}
\caption{Scalefactor for $C=1$ and different values of $Y_0$. The solid (black) curve shows the
usual {\sc FRW} case with a big bang singularity. \label{fig1}}
\end{figure}

In Figure \ref{fig1} we show the dependence of the solution for the scale-factor on the initial value $Y_0$, for a selected
value of $C=1$. One sees, as expected,
that the curves track each other around the present time $t/t_0 = 1$ and diverge away from that.  
The larger $Y_0$ the stronger the divergence. While not clearly visible in the plot, all curves approach
the same (usual) scaling behavior at large $t$. 

\begin{figure}[th]
\centering
\vspace*{-1.3cm}\hspace*{1cm}
\includegraphics[width=10cm]{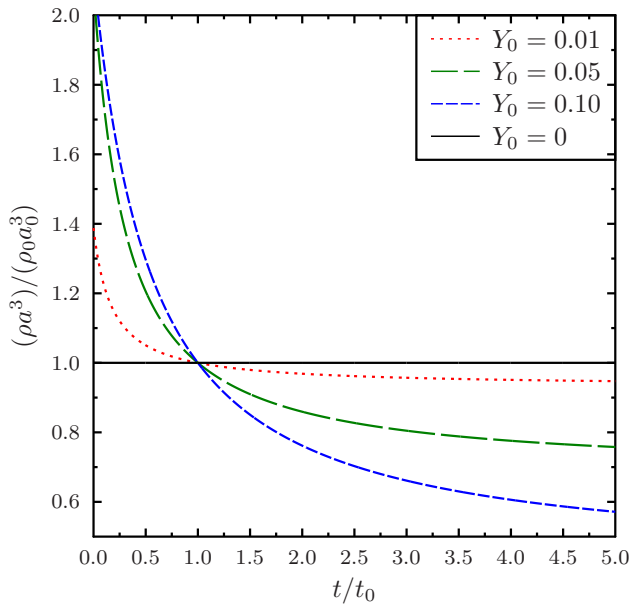}
\vspace*{-1cm}
\hspace*{2cm}\caption{Density times volume for $C=1$ and different values of $Y_0$.  \label{fig2}}
\end{figure}

In Figure \ref{fig2} we plot the matter density multiplied by the 3-volume, 
$\rho a^3$, which for usual
{\sc FRW} case is a constant. We can see that, here too, the solutions closely track each other at present
times. At large times they all become constant. 

The dependence of the curves on $C$ is rather uninteresting, though that in itself is interesting. The
constant $C$ only makes a noticeable difference if it is comparable to or larger than $1/Y_0^2$, and
then it does not lead to qualitatively new behavior but slightly shifts the curves up and down.

\section{Discussion}
\label{disc}

We note that both in Causal Sets as also in Spin-Foam approaches to
quantum gravity, the density of nodes in the network describes the
volume. Since we have seen from Eq.~(\ref{volel}) that the formalism
used here describes a change in the volume-element, it might therefore
also be useful to study fluctuations of the volume-measure around
the mean value.

Further, we note that in the treatment presented here the defects do not couple to gauge
fields. This is because the symmetric part of the connection does not appear in
the field-strength tensor. For this reason, one does not reproduce the
previously considered flat-space
model in which it was assumed that the modified connection also couples
to the gauge fields.

In principle one can look at the general case with non-metricity and torsion.
But before doing so it would make more sense to tighten the relation between
space-time defects and particular changes to the covariant derivative. While
we believe that a stochastic kick to a particle's momentum is a relatively
straight-forward modification of the derivative, it is not a priori clear what
would give rise to a torsion-like contribution.

\section{Conclusion}

The work presented here answers
two questions about space-time defects raised in \cite{Hossenfelder:2013zda}: a) What is the
time-dependence of the average density of local space-time defects in an expanding
universe and b) Does the presence of this average contribution from the space-time defects affect the
expansion of the universe. The answer to question a) is that the correction
term to the derivative drops with $1/a^3$, just like the baryonic density.
The answer to question b) is that the average contribution from space-time defects
to the dynamics of the universe is relevant only at early (Planckian) times and
the effect it has at the present time is negligble relative to that of radiation.

While we have here not taken into account stochastic deviations from the average,
the time-dependence calculated in this present work provides us with the 
mean value necessary to assess possible observational consequences of local
space-time defects. 

\section*{Acknowledgements}

We thank the Foundational Questions Institute FQXi for support.

\end{document}